\documentclass[10pt,conference]{IEEEtran}

\usepackage[dvipdfm]{graphicx}
\usepackage[fleqn]{amsmath}
\usepackage{amssymb}
\usepackage{bm}

\newtheorem{definition}{Definition}
\newtheorem{theorem}{Theorem}
\newtheorem{lemma}{Lemma}

\renewcommand{\QED}{\QEDopen}

\begin{document}

\title{Relations between the Local Weight Distributions of a Linear Block Code, Its Extended Code, and Its Even Weight Subcode}

\author{\authorblockN{Kenji Yasunaga\authorrefmark{1} 
and Toru Fujiwara\authorrefmark{2}}
\authorblockA{Graduate School of Information Science and Technology\\
Osaka University\\
1-5 Yamadaoka, Suita, Osaka 565-0871, Japan\\
E-mail: \{k-yasunaga\authorrefmark{1}, fujiwara\authorrefmark{2}\}@ist.osaka-u.ac.jp}
}

%

\maketitle

\begin{abstract}
Relations between the local weight distributions of a binary linear code,
its extended code, and its even weight subcode are presented.
In particular, for a code of which the extended code is transitive
invariant and contains only codewords with weight multiples of four, 
the local weight distribution can be obtained from that of the extended code.
Using the relations, the local weight distributions of the $(127,k)$ 
primitive BCH codes for $k\leq50$, the $(127,64)$ punctured third-order Reed-Muller , 
and their even weight subcodes are 
obtained from the local weight distribution of the $(128,k)$ extended primitive BCH 
codes for $k\leq50$ and the $(128,64)$ third-order Reed-Muller code.
We also show an approach to improve an algorithm for computing the
local weight distribution proposed before.

\end{abstract}

\section{Introduction}\label{sec:intro}
In a binary linear code, a zero neighbor is a codeword whose Voronoi region shares
a facet with that of the all-zero codeword~\cite{agrell96}.
The local weight distribution~\cite{agrell98, yasunaga04}
(or local distance profile~\cite{agrell96, yasunaga04_2,mohri02, mohri03, forney}) 
of a binary linear code is defined as the weight distribution of zero neighbors 
in the code.
Knowledge of the local weight distribution of a code is valuable for the error 
performance analysis of the code. For example, the local weight distribution could give
a tighter upper bound on error probability for soft decision decoding over AWGN channel
than the usual union bound~\cite{forney}.

Formulas for local weight distribution are only known for certain classes of codes, 
Hamming codes and second-order Reed-Muller codes.
Although an efficient method to examine zero neighborship of codeword is presented
in~\cite{agrell96}, the computation for obtaining the local weight distribution is
a very time-consuming task.
As Agrell noted in~\cite{agrell96}, the automorphism group of the code can help reduce 
the complexity. Using the automorphism group of cyclic codes, i.e. cyclic permutations, 
Mohri et al. devised a computation algorithm using a method for obtaining the 
representative codewords with respect to the cyclic permutations.
The algorithm examines the zero neighborship only for the representative codewords.
By using the algorithm, they obtained the local weight distributions of 
the binary primitive BCH codes of length 63~\cite{mohri02,mohri03}.

When the automorphism group properly contains cyclic permutations,
for example the affine group, an efficient way of finding the representative 
codewords is unknown.
We proposed an algorithm for computing the local weight distribution of a code
using the automorphism group of the set of the all cosets of a subcode in the
code~\cite{yasunaga04_2}.
Using the algorithm, we obtained the local weight distributions of 
the $(128,k)$ extended primitive BCH codes 
for $k\leq50$ and the $(128,64)$ third-order 
Reed-Muller code~\cite{yasunaga04,yasunaga04_2}. For extended primitive BCH codes,
which is closed under the affine group of permutations, our proposed algorithm
has considerably smaller complexity than the algorithm 
in~\cite{mohri02} and~\cite{mohri03}.

However, for cyclic codes, the complexity is not reduced.
Then the local weight distributions of the $(127,k)$ primitive BCH codes for 
$k\geq36$ were not obtained although those of the corresponding $(128,k)$ 
extended primitive BCH codes are known.
A method for obtaining the local weight distribution of a code from that of 
its extended code should be considered.

In this paper, a relation between local weight distributions of a 
binary linear code and its extended code is given.
A more concrete relation is presented for the case that
the extended code is transitive invariant and contains only 
codewords with weight multiples of four.
Extended binary primitive BCH codes and Reed-Muller codes are transitive invariant codes.
A relation between local weight distributions of a 
binary linear code and its even weight subcode is also given.
By using the relations, the local weight distributions of the $(127,k)$ binary 
primitive BCH codes for $36\leq k \leq50$, the $(127,64)$ punctured third-order 
Reed-Muller code, and their even weight subcodes 
are obtained from the local weight distributions of the $(128,k)$ primitive BCH
codes for $36\leq k \leq50$ and the $(128,64)$ third-order Reed-Muller code.
Finally, we give an approach to improve the algorithm proposed in~\cite{yasunaga04_2}.

\section{Local Weight Distribution}

Let $C$ be a binary $(n,k)$ linear code. 
Define a mapping $s$ from $\{0,1\}$ to $\mathbf{R}$ 
as $s(0)=1$ and $s(1)=-1$.
The mapping $s$ is naturally extended 
to one from $\{0,1\}^n$ to $\mathbf{R}^n$.
A zero neighbor of $C$ is defined~\cite{agrell96} as follows:
\medskip
\begin{definition}[Zero neighbor]\label{def:zn}
For $\bm{v} \in C$, define $\bm{m}_0\in \mathbf{R}^n$ as 
$\bm{m}_0=\frac{1}{2}(s(\bm{0})+s(\bm{v}))$
where $\bm{0}=(0,0,\ldots,0)$. The codeword $\bm{v}$ is a zero neighbor if and only if
\begin{eqnarray}
&&d_E(\bm{m}_0,s(\bm{v}))=d_E(\bm{m}_0,s(\bm{0}))<d_E(\bm{m}_0,s(\bm{v}')),
\nonumber\\
&&\hspace*{30mm}\mbox{for any}\ \bm{v}'\in C\setminus \{ \bm{0},\bm{v} \},
\end{eqnarray}
where $d_E$($\bm{x}, \bm{y}$) is the Euclidean distance between $\bm{x}$ 
and $\bm{y}$ in $\mathbf{R}^n$.
\end{definition}
\medskip

A zero neighbor is also called a minimal codeword in~\cite{ashikhmin98}.
The following lemma is useful to check whether a given codeword is 
a zero neighbor or not~\cite{agrell96}.
\medskip
\begin{lemma}\label{lem:zn} 
$\bm{v} \in C$ is a zero neighbor if and only if there is not $\bm{v}'$
$\in C\setminus \{\bm{0}\}$ such that ${\rm Supp}(\bm{v}') \subsetneq 
{\rm Supp}(\bm{v})$.
Note that ${\rm Supp}(\bm{v})$ is the set of support of $\bm{v}$, 
which is the set of positions of nonzero elements in $\bm{v} = (v_1,v_2,\ldots,v_n)$.
\end{lemma}
\medskip

The local weight distribution is defined as follows:
\medskip
\begin{definition}[Local weight distribution]\label{def:ldp}
Let $L_w(C)$ be the number of zero neighbors with weight $w$ in $C$. 
The local weight distribution of $C$ is defined as the 
$(n+1)$-tuple $(L_0(C), L_1(C), \ldots, L_n(C))$.
\end{definition}
\medskip

On the local weight distribution, we have the following 
lemma~\cite{agrell98,ashikhmin98}.

\medskip
\begin{lemma}\label{lem:condition}
Let $A_w(C)$ be the number of codewords with weight $w$ in $C$ and
 $d$ be the minimum distance of $C$. 
\begin{equation}
L_w(C)  =  \left\{
\begin{array}{ll}
A_w(C), & w < 2d, \\
0, & w>n-k+1.
\end{array}\right.
\end{equation}
\end{lemma}
\medskip

When all the weights $w$ in a code are confined in $w<2d$ and $w>n-k+1$, the 
local weight distribution can be obtained from the weight distribution straightforwardly.
For example, the local weight distribution of the $(n, k)$ primitive BCH code 
of length 63 for $k\leq18$, of length 127 for $k\leq29$, and of length 255 for $k\leq45$ 
can be obtained from their weight distributions.

\section{Relations of Local Weight Distributions}
\subsection{General relation}
Consider a binary linear code $C$ of length $n$, its extended code $C_{\rm ex}$,
and its even weight subcode $C_{\rm even}$.
For a codeword $\bm{v} \in C$, let ${\rm wt}(\bm{v})$ be the Hamming weight of 
$\bm{v}$ and $\bm{v}^{\rm (ex)}$ be the corresponding codeword in $C_{\rm ex}$, that is, 
$\bm{v}^{\rm (ex)}$ is obtained from $\bm{v}$ by adding the over-all parity bit.
We define a {\it decomposable} codeword.

\begin{figure}[t]
\begin{center}
\includegraphics[width=7cm,clip]{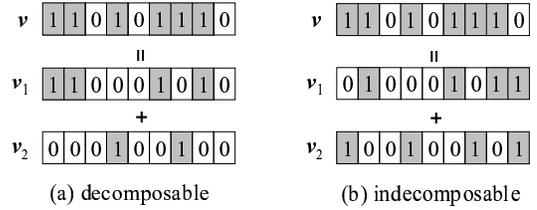}
\end{center}
\caption{Examples of a decomposable codeword and an indecomposable codeword.}
\label{fig:deco}
\end{figure}

\medskip
\begin{definition}[Decomposable codeword]
$\bm{v} \in C$ is called {\it decomposable} if $\bm{v}$ can be represented as 
$\bm{v}=\bm{v}_1+\bm{v}_2$ where $\bm{v}_1, \bm{v}_2 \in C$ and 
${\rm Supp}(\bm{v}_1) \cap {\rm Supp}(\bm{v}_2) = \emptyset$ (see Figure~\ref{fig:deco}).
\end{definition}
\medskip
From Lemma~\ref{lem:zn}, $\bm{v}$ is not a zero neighbor if and only if 
$\bm{v}$ is decomposable. For even weight codewords, 
we introduce an {\it only-odd-decomposable} codeword and an {\it even-decomposable}
codeword.

\medskip
\begin{definition}
Let $\bm{v} \in C$ be a decomposable codeword with even ${\rm wt}(\bm{v})$.
That is, $\bm{v}$ is not a zero neighbor in $C$.
$\bm{v}$ is said to be {\it only-odd-decomposable}, if all the decomposition of $\bm{v}$
is of the form $\bm{v}_1+\bm{v}_2$ with the odd weight codewords
$\bm{v}_1, \bm{v}_2 \in C$. Otherwise, $\bm{v}$ is said to be {\it even-decomposable}.
\end{definition}
\medskip
When $\bm{v}$ is even-decomposable, there is a decomposition of $\bm{v}$,
$\bm{v}_1+\bm{v}_2$ such that both ${\rm wt}(\bm{v}_1)$ and ${\rm wt}(\bm{v}_2)$
are even.

The relation between $C$ and $C_{\rm ex}$ with respect to zero neighborship 
is given in the following theorem, which is also summarized in Table~\ref{tb:zn}.

\begin{figure}[t]
\begin{center}
\includegraphics[width=8.5cm,clip]{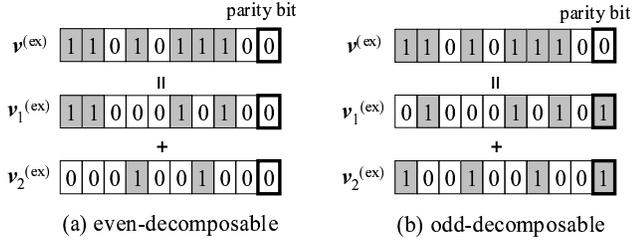}
\end{center}
\caption{Examples of an even-decomposable codeword and an odd-decomposable codeword mentioned in the proof of Theorem~\ref{th:ori_ex}-2)-b).}
\label{fig:even_odd_deco}
\end{figure}

\medskip
\begin{theorem}\normalfont\label{th:ori_ex}
\begin{enumerate}
\item For a zero neighbor $\bm{v}$ in $C$, $\bm{v}^{\rm (ex)}$ is 
a zero neighbor in $C_{\rm ex}$.
\item For a codeword $\bm{v}$ which is not a zero neighbor in $C$,
the following a) and b) hold.
\begin{enumerate}
\item When ${\rm wt}(\bm{v})$ is odd, $\bm{v}^{\rm (ex)}$ is 
not a zero neighbor in $C_{\rm ex}$.
\item When ${\rm wt}(\bm{v})$ is even, $\bm{v}^{\rm (ex)}$ is a zero neighbor 
in $C_{\rm ex}$ if and only if $\bm{v}$ is only-odd-decomposable in $C$.
\end{enumerate}
\end{enumerate}
\end{theorem}
\medskip
\begin{proof}
1) Suppose that $\bm{v}^{\rm (ex)}$ is not a zero neighbor in $C_{\rm ex}$.
Then $\bm{v}^{\rm (ex)}$ is decomposable into 
$\bm{v}_1^{\rm (ex)}+\bm{v}_2^{\rm (ex)}$, and hence $\bm{v}$ is decomposable into
$\bm{v}_1+\bm{v}_2$. This contradicts the indecomposability of $\bm{v}$.

2) Suppose that $\bm{v}$ is decomposed into $\bm{v}=\bm{v}_1+\bm{v}_2$.
a) Since ${\rm wt}(\bm{v})$ is odd, the sum of the parity bits in 
$\bm{v}_1^{\rm (ex)}$ and $\bm{v}_2^{\rm (ex)}$ is one.
Also, the parity bit in $\bm{v}^{\rm (ex)}$ is one.
Then $\bm{v}^{\rm (ex)}$ is decomposable into
$\bm{v}_1^{\rm (ex)}+\bm{v}_2^{\rm (ex)}$, and $\bm{v}^{\rm (ex)}$
is not a zero neighbor in $C_{\rm ex}$.
b) Since ${\rm wt}(\bm{v})$ is even, the parity bit in $\bm{v}^{\rm (ex)}$ is zero.
(If part) Suppose that $\bm{v}^{\rm (ex)}$ is not a zero neighbor in $C_{\rm ex}$.
Then there exists a decomposition 
$\bm{v}^{\rm (ex)}=\bm{v}_1^{\rm (ex)}+\bm{v}_2^{\rm (ex)}$.
Because the parity bit in $\bm{v}^{\rm (ex)}$ is zero, the parity bits in
$\bm{v}_1^{\rm (ex)}$ and $\bm{v}_2^{\rm (ex)}$ must be zero.
This implies that $\bm{v}$ is even-decomposable into $\bm{v}_1+\bm{v}_2$,
and contradicts the assumption that $\bm{v}$ is only-odd-decomposable.
(Only if part) Suppose that $\bm{v}$ is even-decomposable.
Then there is a decomposition such that
the parity bits in both $\bm{v}_1^{\rm (ex)}$ and $\bm{v}_2^{\rm (ex)}$ are zero.
For such the decomposition, $\bm{v}^{\rm (ex)}$ is decomposable into 
$\bm{v}_1^{\rm (ex)}+\bm{v}_2^{\rm (ex)}$, and $\bm{v}^{\rm (ex)}$
is not a zero neighbor in $C_{\rm ex}$ (see Figure~\ref{fig:even_odd_deco}).
\end{proof}
\medskip

From 2)-b) of Theorem~\ref{th:ori_ex}, 
there may be codewords that are not zero neighbors in $C$
although their extended codewords are zero neighbors in $C_{\rm ex}$.
Such codewords are the only-odd decomposable codewords.
For investigating relations of local weight distributions between a code and its
extended code, only-odd decomposable codewords are important.

The following theorem is a direct consequence of Theorem~\ref{th:ori_ex}.
\medskip
\begin{theorem}\label{th:ori_ex_equ}
For a code $C$ of length $n$,
\begin{equation}
L_{2i}(C_{\rm ex})  = L_{2i-1}(C)+L_{2i}(C)+N_{2i}(C), \ 0 \leq i \leq n/2,
\end{equation}
where $N_{j}(C)$ is the number of only-odd decomposable codewords with weight $j$
in $C$.
\end{theorem}
\medskip

\begin{table*}[t]
\caption{Zero neighborship of $\bm{v}$ in a linear block code, $\bm{v}_{\rm ex}$
in its extended code, and $\bm{v}$ in its even weight subcode.}\label{tb:zn}
\begin{center}
\begin{tabular}{|c|c|c|c|c|c|c|} \hline
\multicolumn{3}{|c|}{$\bm{v}$ in $C$} & \multicolumn{2}{|c|}{$\bm{v}^{\rm (ex)}$ in $C_{\rm ex}$} & \multicolumn{2}{|c|}{$\bm{v}$ in $C_{\rm even}$}\\ \hline
Zero neighborship & Weight & Decomposability & Zero neighborship & Theorem~\ref{th:ori_ex} & Zero neighborship & Theorem~\ref{th:ori_even} \\\hline\hline
     & Odd  &  &  & & N/A & N/A \\ \cline{2-2} \cline{6-7}
\raisebox{1ex}[0pt]{Yes} & Even & \raisebox{1ex}[0pt]{Not decomposable} & \raisebox{1ex}[0pt]{Yes}  & \raisebox{1ex}[0pt]{1)} & Yes & 1) \\ \hline
 & Odd  & Decomposable & No & 2) - a) & N/A & N/A \\ \cline{2-7}
No  & Even & Only-odd-decomposable & Yes & & Yes & \\ \cline{2-4}\cline{6-6}
 & Even & Even-decomposable     & No  & \raisebox{1ex}[0pt]{2) - b)} &  No & \raisebox{1ex}[0pt]{2)}  \\ \hline
\end{tabular}
\end{center}
\end{table*}

From Theorem~\ref{th:ori_ex_equ}, 
if there is no only-odd decomposable codeword in $C$,
the local weight distributions of $C_{\rm ex}$ are obtained from that of $C$. 
Next, we give a useful sufficient condition 
under which no only-odd-decomposable codeword exists.

\medskip
\begin{theorem}\label{th:multi_four}
If all the weights of codewords in $C_{\rm ex}$ are multiples of four, 
no only-odd-decomposable codeword exists in $C$.
\end{theorem}
\medskip
\begin{proof}
If $\bm{v} \in C$ is an only-odd-decomposable codeword and decomposed into
$\bm{v}_1+\bm{v}_2$,
the weights of $\bm{v}_1$ and $\bm{v}_2$ can be represented as
 ${\rm wt}(\bm{v}_1) = 4i-1$ and ${\rm wt}(\bm{v}_2) = 4j-1$ where $i$ and $j$ 
are integers. Then ${\rm wt}(\bm{v}) = {\rm wt}(\bm{v}_1+\bm{v}_2)= 
{\rm wt}(\bm{v}_1)+{\rm wt}(\bm{v}_2) = (4i-1)+(4j-1)= 4i+4j-2$. 
This contradicts the fact that ${\rm wt}(\bm{v})$ is a multiple of four.
\end{proof}
\medskip

For example, all the weights of codewords in the $(128,k)$ extended primitive BCH 
code with $k\leq57$ are multiples of four. 
The parameters of Reed-Muller codes with which all the weights of codewords 
are multiples of four are given by Corollary 13 of Chapter 15 in~\cite{williams}.
From the corollary, the third-order Reed-Muller codes of length 128, 256, and 512 
have only codewords whose weights are multiples of four.

Although the local weight distribution of $C_{\rm ex}$ 
for these codes can be obtained from that of $C$ by using 
Theorem~\ref{th:ori_ex_equ}, what we need is 
a method for obtaining the local weight distribution of $C$ from that of $C_{\rm ex}$.
We need to know the number of zero neighbors with parity bit one.
In Section~\ref{sec:rel_tran}, we will show a method to obtain the number of 
zero neighbors with parity bit one for a class of transitive invariant codes.

A similar relation to that between $C$ and $C_{\rm ex}$ 
holds between $C$ and $C_{\rm even}$.
This relation is given in Theorem~\ref{th:ori_even} 
without proof (see Table~\ref{tb:zn}).

\medskip

\begin{theorem}\label{th:ori_even}
\begin{enumerate}
\item For an even weight zero neighbor $\bm{v}$ in $C$,
$\bm{v}$ is a zero neighbor in $C_{\rm even}$.
\item For an even weight codeword $\bm{v}$ which is not a zero neighbor in $C$,
$\bm{v}$ is a zero neighbor in $C_{\rm even}$ 
if and only if $\bm{v}$ is only-odd-decomposable in $C$.
\end{enumerate}
\end{theorem}

\medskip
From Theorem~\ref{th:ori_even}, 
we have Theorem~\ref{th:ori_even_equ}.
\medskip

\begin{theorem}\label{th:ori_even_equ}
For a code $C$ of length $n$,
\begin{eqnarray} 
L_{2i}(C_{\rm even}) & = & L_{2i}(C)+N_{2i}(C),  \ \ \ \ 
0\leq i \leq n/2.
\end{eqnarray}
\end{theorem}
\medskip

\subsection{Relation for transitive invariant extended codes}\label{sec:rel_tran}
A transitive invariant code is a code which is invariant under 
a transitive group of permutations.
A group of permutations is said to be transitive if for any two symbols in a codeword
there exists a permutation that interchanges them~\cite{peterson}.
The extended primitive BCH codes and Reed-Muller codes are transitive invariant codes.
For a transitive invariant $C_{\rm ex}$, a relation between the (global) weight 
distributions of $C$ and $C_{\rm ex}$ is presented in Theorem 8.15 in~\cite{peterson}. 
A similar relation holds for local weight distribution.

\medskip
\begin{lemma}\label{lem:zn2}
If $C_{\rm ex}$ is a transitive invariant code of length $n+1$, 
the number of zero neighbors with parity bit one is $\frac{w}{n+1} L_{w}(C_{\rm ex})$.
\end{lemma}
\medskip
\begin{proof}
This lemma can be proved in a similar way as the proof of Theorem 8.15.
Arrange all zero neighbors with weight $w$ in a column. Next, interchange
the $j$-th column and the last column, which is the parity bit column,
for all these codewords with the permutation. All the resulting codewords
have weight $w$ and must be the same as the original set of codewords. 
Thus, the number of ones in the $j$-th column and that in the last column are the same.
Denote this number $l_w$, which is the same as the number of zero neighbors of
weight $w$ with parity bit one.
Then the total ones in the original set of codewords is $(n+1) \, l_w$, 
or  $L_w(C_{\rm ex})$ times the weight $w$. Thus, $(n+1)\, l_w=wL_w(C_{\rm ex})$, and
$l_w = \frac{w}{n+1} L_{w}(C_{\rm ex})$.
\end{proof}
\medskip

It is clear that there are $\frac{n+1-w}{n+1} L_w(C_{\rm ex})$ zero neighbors with 
weight $w$ whose parity bit is zero from this lemma.
The following theorem is obtained from Theorem~\ref{th:ori_ex} and 
Lemma~\ref{lem:zn2}.

\medskip
\begin{theorem}\label{th:transitive}
If $C_{\rm ex}$ is a transitive invariant code of length $n+1$,
\begin{eqnarray}
L_w(C) & = & \frac{w+1}{n+1} L_{w+1}(C_{\rm ex}), \ \ \ \ \ \ 
{\rm for} \ \ {\rm odd} \ \ w,\label{eq:transitive_eq}\\
L_w(C) & = & \frac{n+1-w}{n} L_w(C_{\rm ex}) - N_w(C),\\
& \leq & \frac{n+1-w}{n+1} L_w(C_{\rm ex}),\ \ \ \,
{\rm for}\ \ {\rm even}\ \ w.\label{eq:transitive_leq}
\end{eqnarray}
If there is no only-odd-decomposable codeword in a transitive invariant code
$C$, the equality of~(\ref{eq:transitive_leq}) holds. That is, in this case,
we have that
\begin{eqnarray}
L_w(C) & = & \frac{n+1-w}{n} L_w(C_{\rm ex}),\ \ \ 
{\rm for}\ \ {\rm even}\ \ w .\label{eq:transitive_eq2}
\end{eqnarray}
\end{theorem}
\bigskip

Therefore, for a transitive invariant code $C_{\rm ex}$ having no 
only-odd-decomposable codeword in $C$, the local weight distributions of 
$C$  can be obtained from that of $C_{\rm ex}$ by using 
(\ref{eq:transitive_eq}) and (\ref{eq:transitive_eq2}) in Theorem~\ref{th:transitive}.
After computing the local weight distribution of $C$, that of $C_{\rm even}$
can be obtained by using Theorem~\ref{th:ori_even_equ}.

\section{Obtained Local Weight Distributions}
As discussed in the previous section, the local weight distributions of the $(127,k)$
primitive BCH codes for $k\leq57$, the punctured third-order Reed-Muller codes 
of length 127, 255, and 511, and their even weight subcodes are obtained from 
those of the corresponding extended codes by using
Theorems~\ref{th:ori_even_equ} and~\ref{th:transitive}.
Since the local weight distribution for the $(128,57)$ extended primitive BCH code 
and the third-order Reed-Muller codes of length 256 and 512 are unknown, 
only the local weight distributions of the $(127,k)$ primitive BCH codes 
for $k=36,43,50$, the $(127,64)$ punctured third-order Reed-Muller code,
and their even weight subcodes are obtained.
These local weight distributions are presented in Table~\ref{tb:lwd_bch}.
The local weight distributions of the corresponding even weight subcodes are
obtained straightforwardly from the distributions given in the table by
using Theorem~\ref{th:ori_even_equ}.

\begin{table*}[htbp]
\caption{The local weight distributions of the $(127,k)$ primitive BCH codes
for $k=36,43, {\rm and} \ 50$ and the punctured third-order Reed-Muller code of length 127 .}\label{tb:lwd_bch}
\begin{center}
\begin{tabular}{|c|r|c|c|r|c|c|r|c|c|r|}
\multicolumn{2}{c}{$(127,36)$ BCH code} & \multicolumn{1}{c}{} & \multicolumn{2}{c}{$(127,43)$ BCH code} & \multicolumn{1}{c}{} & \multicolumn{2}{c}{$(127,50)$ BCH code} & \multicolumn{1}{c}{} & \multicolumn{2}{c}{$(127,64)$ punc. RM code} \\ \cline{1-2} \cline{4-5} \cline{7-8} \cline{10-11}
$w$ & \multicolumn{1}{|c|}{$L_w$} & & $w$ & \multicolumn{1}{|c|}{$L_w$} & & $w$ & \multicolumn{1}{|c|}{$L_w$} & & $w$ & \multicolumn{1}{|c|}{$L_w$} \\ \cline{1-2} \cline{4-5} \cline{7-8} \cline{10-11}
31 & 2,667 & & 31 & 31,115 & & 27 & 40,894 & & 15 & 11,811 \\ \cline{1-2} \cline{4-5} \cline{7-8} \cline{10-11}
32 & 8,001 & & 32 & 93,345 & & 28 & 146,050 & & 16 &  82,677 \\ \cline{1-2} \cline{4-5} \cline{7-8} \cline{10-11}
35 & 4,572 & & 35 & 2,478,024 & & 31 & 4,853,051 & & 23 & 13,889,736  \\ \cline{1-2} \cline{4-5} \cline{7-8} \cline{10-11}
36 & 11,684 & & 36 & 6,332,728 & & 32 & 14,559,153 & & 24 & 60,188,856  \\ \cline{1-2} \cline{4-5} \cline{7-8} \cline{10-11}
39 & 640,080 & & 39 & 82,356,960 & & 35 & 310,454,802  & & 27 & 684,345,088 \\ \cline{1-2} \cline{4-5} \cline{7-8} \cline{10-11}
40 & 1,408,176 & & 40 & 181,185,312 & & 36 & 793,384,494  & & 28 & 2,444,089,600 \\ \cline{1-2} \cline{4-5} \cline{7-8} \cline{10-11}
43 & 12,220,956 & & 43 & 1,554,145,736 & & 39 & 10,538,703,840 & & 31 &  77,893,639,488 \\ \cline{1-2} \cline{4-5} \cline{7-8} \cline{10-11}
44 & 23,330,916 & & 44 & 2,967,005,496 & & 40 & 23,185,148,448 & & 32 &  233,680,918,464 \\ \cline{1-2} \cline{4-5} \cline{7-8} \cline{10-11}
47 & 132,560,568 & & 47 & 16,837,453,752 & & 43 & 199,123,183,160 & & 35 &  5,097,898,213,632 \\ \cline{1-2} \cline{4-5} \cline{7-8} \cline{10-11}
48 & 220,934,280 & & 48 & 28,062,422,920 & & 44 & 380,144,258,760 & & 36 & 13,027,962,101,504\\ \cline{1-2} \cline{4-5} \cline{7-8} \cline{10-11}
51 & 823,921,644 & & 51 & 106,485,735,720 & & 47 & 2,154,195,406,104 & & 39 & 172,489,249,981,440\\ \cline{1-2} \cline{4-5} \cline{7-8} \cline{10-11}
52 & 1,204,193,172 & & 52 & 155,632,998,360 & & 48 & 3,590,325,676,840 & & 40 & 379,476,349,959,168\\ \cline{1-2} \cline{4-5} \cline{7-8} \cline{10-11}
55 & 3,157,059,472 & & 55 & 400,716,792,672 & & 51 & 13,633,106,229,288 & & 43 & 3,259,718,804,643,840\\ \cline{1-2} \cline{4-5} \cline{7-8} \cline{10-11}
56 & 4,059,076,464 & & 56 & 515,207,304,864 & & 52 & 19,925,309,104,344 & & 44 & 6,223,099,536,138,240\\ \cline{1-2} \cline{4-5} \cline{7-8} \cline{10-11}
59 & 7,022,797,740 & & 59 & 905,612,814,120 & & 55 & 51,285,782,220,204 & & 47 & 35,130,035,853,803,520\\ \cline{1-2} \cline{4-5} \cline{7-8} \cline{10-11}
60 & 7,959,170,772 & & 60 & 1,026,361,189,336 & & 56 & 65,938,862,854,548 & & 48 & 58,550,059,756,339,200\\ \cline{1-2} \cline{4-5} \cline{7-8} \cline{10-11}
63 & 9,742,066,368 & & 63 & 1,238,334,929,472 & & 59 & 115,927,157,830,260 & & 51 & 218,602,288,622,075,904\\ \cline{1-2} \cline{4-5} \cline{7-8} \cline{10-11}
64 & 9,742,066,368 & & 64 & 1,238,334,929,472 & & 60 & 131,384,112,207,628 & & 52 & 319,495,652,601,495,552\\ \cline{1-2} \cline{4-5} \cline{7-8} \cline{10-11}
67 & 7,959,170,772 & & 67 & 1,026,345,592,720 & & 63 & 158,486,906,385,472 & & 55 & 766,899,891,905,495,040\\ \cline{1-2} \cline{4-5} \cline{7-8} \cline{10-11}
68 & 7,022,797,740 & & 68 & 905,599,052,400 & & 64 & 158,486,906,385,472 & & 56 & 986,014,146,735,636,480\\ \cline{1-2} \cline{4-5} \cline{7-8} \cline{10-11}
71 & 4,059,071,892 & & 71 & 515,097,101,376 & & 67 & 131,258,388,369,668 & & 59 & 1,306,771,964,441,395,200\\ \cline{1-2} \cline{4-5} \cline{7-8} \cline{10-11}
72 & 3,157,055,916 & & 72 & 400,631,078,848 & & 68 & 115,816,225,032,060 & & 60 & 1,481,008,226,366,914,560\\ \cline{1-2} \cline{4-5} \cline{7-8} \cline{10-11}
75 & 1,204,193,172 & & 75 & 155,191,535,184 & & 71 & 64,917,266,933,304 & & 63 & 258,664,522,171,023,360\\ \cline{1-2} \cline{4-5} \cline{7-8} \cline{10-11}
76 & 823,921,644 & & 76 & 106,183,681,968 & & 72 & 50,491,207,614,792 & & 64 & 258,664,522,171,023,360\\ \cline{1-2} \cline{4-5} \cline{7-8} \cline{10-11}
79 & 217,627,200 & & 79 & 26,980,367,680 & & 75 & 15,345,182,164,032 & \multicolumn{3}{c}{} \\ \cline{1-2} \cline{4-5} \cline{7-8}
80 & 130,576,320 & & 80 & 16,188,220,608 & & 76 & 10,499,335,164,864 & \multicolumn{3}{c}{} \\ \cline{1-2} \cline{4-5} \cline{7-8}
83 & 23,330,916 & & 83 & 1,617,588,840 & \multicolumn{6}{c}{}  \\ \cline{1-2} \cline{4-5} 
84 & 12,220,956 & & 84 & 847,308,440 & \multicolumn{6}{c}{}  \\ \cline{1-2} \cline{4-5}
87 & 1,408,176 & \multicolumn{8}{c}{} \\ \cline{1-2} 
88 & 640,080 &  \multicolumn{8}{c}{}  \\ \cline{1-2} 
\end{tabular}
\end{center}
\end{table*}

\section{Computation Algorithm for Longer Codes}
As mentioned in Section~\ref{sec:intro}, we have computed the local weight distributions
of several extended BCH codes and the third-order Reed-Muller code of length 128 by
using our computation algorithm in~\cite{yasunaga04_2}.
In the algorithm, the time complexity is reduced by reducing the number of codewords
whose zero neighborship should be checked.
However, it is still time consuming to compute the local weight distribution of a
$(n,k)$ code with larger $k$ and/or larger $n$.
In improving the computation algorithm, the next target codes include
the $(256,93)$ Reed-Muller code and the $(128,57)$ extended BCH code.

In this section, we give an approach for improving 
the algorithm proposed in~\cite{yasunaga04_2}.
First, we briefly review this algorithm.
For a binary $(n,k)$ linear code $C$ and its linear subcode with dimension $k'$,
let $C/C'$ denote the set of cosets of $C'$ in $C$, that is,
$C/C' = \{ \bm{v} + C': \bm{v} \in C\setminus C' \}$.
Then,
\begin{equation}
|C/C'| =2^{k-k'},\quad\mbox{and}\quad C = \bigcup_{D\in C/C'}D.
\label{eqn:coset}
\end{equation}

\medskip
\begin{definition}[Local weight subdistribution for cosets]
Let $D$ be a coset in $C/C'$ and $LS_w(D)$ be 
the number of zero neighbors of $C$ in $D$ with weight $w$.
The local weight subdistribution for a coset $D \in C/C'$ is 
the $(n+1)$-tuple $(LS_0(D),LS_1(D),\ldots,LS_n(D))$.

Then 
\begin{equation}
 L_w(C) = \sum_{D \in C/C'}LS_w(D) .
\end{equation}
\end{definition}
\medskip
The following theorem gives an invariance property under permutations for cosets.
\medskip
\begin{theorem}[Invariance property for cosets]\label{th:inv_coset}
For $D_1,D_2 \in C/C'$, the local weight subdistribution for $D_1$ and 
that for $D_2$ are the same if there exists $\pi \in {\rm Aut}(C)$
such that $\pi[D_1] = D_2$, where ${\rm Aut}(C)$ is the automorphism group of $C$ and 
$\pi[D_1]=\{ \pi\bm{v} : \bm{v} \in D_1 \}$.
\end{theorem}
\medskip
In the algorithm in~\cite{yasunaga04_2}, given a subgroup $P$ of ${\rm Aut}(C)$,
cosets in $C/C'$ are partitioned into
equivalence classes such that two cosets $D_1$ and $D_2$ are in the same class 
if and only if there is $\pi \in P$ with $\pi[D_1]=D_2$.
Then if the local weight subdistributions only for the representative cosets of 
the classes and the sizes of the classes are obtained, 
we could obtain the local weight distribution of $C$. 


Next, we show the approach.
To reduce the complexity more, we consider using the invariance property for cosets
in computing the local weight subdistributions for the representative cosets.
This means that we consider a coset $\bm{v}+C' \in C/C'$ as the set of cosets of $C''$,
where 
$C''$ is a linear subcode of $C'$.

For a coset $\bm{v}+C' \in C/C'$, let $(\bm{v}+C')/C''$ denote the set of all cosets 
of $C''$ in $\bm{v}+C'$, that is, 
$(\bm{v}+C')/C'' = \{ \bm{v} + \bm{u} + C'': \bm{u}\in C'\setminus C'' \}$.
We also call the weight distribution of zero neighbors in $E \in (\bm{v}+C')/C''$
the local weight subdistribution for $E$.
The following theorem gives an invariance property for cosets in $(\bm{v}+C')/C''$.

\medskip
\begin{theorem}
For $E_1,E_2 \in (\bm{v}+C')/C''$, the local weight subdistribution for $E_1$ and 
that for $E_2$ are the same if there exists $\pi \in \{ \rho : \rho\bm{v} \in \bm{v}+C',
\rho \in {\rm Aut}(C)\cap {\rm Aut}(C') \}$ such that $\pi[E_1] = E_2$,
where $\pi[E] = \{ \pi \bm{v} \, : \, \bm{v} \in E \}$.
\end{theorem}
\medskip

We consider partitioning $(\bm{v}+C')/C''$ into the equivalence classes.
Permutations which are used to partition cosets into equivalence classes are
presented in the following lemma.

\medskip
\begin{lemma}\label{lem:coset_equi}
For a coset $\bm{v}+C' \in (\bm{v}+C')/C''$,
\begin{eqnarray}
\lefteqn{\{ \pi: \pi[E] \in (\bm{v}+C')/C''\ {\rm for \ any} \ E \in (\bm{v}+C')/C'' \}} 
\nonumber \\
&=& \hspace{-2.5mm}\{ \rho : \rho \bm{v} \in \bm{v}+C', \
 \rho \in {\rm Aut}(C)\cap{\rm Aut}(C')\cap{\rm Aut}(C'')\}.\nonumber
\end{eqnarray}
\end{lemma}
\medskip
In order to partition cosets into equivalence classes,
we should use permutations presented in  Lemma~\ref{lem:coset_equi}.
Even if ${\rm Aut}(C)$, ${\rm Aut}(C')$, and ${\rm Aut}(C'')$ are known,
we should obtain permutations $\pi$ that satisfy $\pi\bm{v} \in \bm{v}+C'$.

Let ${\rm RM}(r,m)$ denote the $r$-th order Reed-Muller code of length $2^m$.
We consider the case of the $(256,93)$ third-order Reed-Muller code,
denoted by ${\rm RM}(3,8)$.
The equivalent cosets in ${\rm RM}(3,8)/{\rm RM}(2,8)$ are presented in~\cite{hou96},
and there are 32 equivalence classes.
We choose ${\rm RM}(1,8)$ as a subcode of ${\rm RM}(2,8)$.
Then the general affine group~\cite{williams} is a subgroup of 
${\rm Aut}({\rm RM}(3,8)) \cap {\rm Aut}({\rm RM}(2,8)) \cap {\rm Aut}({\rm RM}(1,8))$.
For each cosets in ${\rm RM}(3,8)/{\rm RM}(2,8)$, the estimated time for 
computing the local weight 
subdistribution is about 54 days with the algorithm in~\cite{yasunaga04_2}.
The total estimated time is about 1700 days.
To compute the local weight distribution of ${\rm RM}(3,8)$ in practical time,
we should find the permutations $\pi$ that satisfy $\pi\bm{v} \in \bm{v}+{\rm RM}(2,8)$
for each coset $\bm{v}+{\rm RM}(2,8)$ in ${\rm RM}(3,8)/{\rm RM}(2,8)$.
If we could find more than 50 such permutations for each cosets, 
the local weight distribution of ${\rm RM}(3,8)$ may be computable.

\section{Conclusion}
In this paper, some relations between local weight distributions of a binary linear code,
its extended code, and its even weight subcode are presented.
The local weight distributions of the $(127,k)$ primitive BCH codes for $k=36,43,50$, 
the $(127,64)$ punctured third-order Reed-Muller code, and 
their even weight subcodes are obtained. 
If the local weight distribution of the $(128,57)$ extended primitive BCH
code and the $(256,93)$ third-order Reed-Muller code are obtained, we can obtain 
the local weight distributions of the $(127,57)$ primitive BCH code, 
the $(255,93)$ punctured third-order Reed-Muller code, and their even weight subcodes.

\end{document}